\documentclass[12pt,preprint]{aastex}

\shorttitle{Lens B1608+656: V, I, and H-band HST Imaging}
\shortauthors{Surpi \& Blandford}

\begin{document}
\date{}  
\title{THE GRAVITATIONAL LENS B1608+656: \\ I. V, I, AND H-BAND HST IMAGING}
\author{G. SURPI\altaffilmark{1} AND R. D. BLANDFORD\altaffilmark{2}}
\affil{California Institute of Technology, 130-33, Pasadena, CA 91125}
\altaffiltext{1}{surpi@tapir.caltech.edu}
\altaffiltext{2}{rdb@tapir.caltech.edu}

\begin{abstract}

We present a multi-wavelength analysis of high-resolution observations 
of the quadruple lens B1608+656 from the HST archive,
acquired with WFPC2 through filters 
F606W (V-band) and F814W (I-band), and with NIC1
in filter F160W (H-band). 

In the three bands, the observations show extended emission 
from the four images of the 
source in a ring-like 
configuration that surrounds the two, resolved, lensing galaxies. 
B1608+656 was discovered as a double-lobed radio source, and later
identified as a post-starburst galaxy in the optical.
Based on photometry and optical spectroscopy
we estimate that the stellar population of the source has an age of $\sim$ 500 Myr.
This provides a model for the spectrum of the source 
that extends over spectral regions where no observations are available,
and is used to generate Tiny Tim PSFs for the filters.
Deconvolutions performed with the Lucy-Richardson method
are presented,
and the limitations of these restorations is discussed. 
V$-$I and I$-$H color maps show evidence of extinction by dust
associated with one of the lensing galaxies, a late type galaxy
presumably disrupted after its close encounter
with the other lens, an elliptical galaxy. 
The extinction affects the two lens galaxies and two of the four 
multiple images. The diagnostic of wavelength-dependent
effects in the images shows that corrections for contamination with 
light from the lenses, extinction, and PSF convolution 
need to be applied before using the extended structure 
in the images as a constraint on lens models.
We will present the restoration of the images in a subsequent paper.
 
\end{abstract}


\section{INTRODUCTION}
\label{intro}

The discovery of the first gravitational lens was announced by 
\citet{wal79}, long after a number of theoretical papers
on lensing had appeared in the 1930s and 1960s. 
Since then several consequences of the lensing effect,
including strong lensing, weak lensing, and microlensing, have been 
observed. In a strong lensing situation,
a gravitational deflector lying close to the line of sight 
to a background source creates multiple images of the source.
Strong lenses provide a powerful tool to study the distribution
of matter, including the dark matter, in the lensing galaxies,
and to measure the Hubble constant
from the time delay between the multiple images \citep{ref64}.  
One of the most promising candidates for this analysis
is the quadruple lens B1608+656,  
due to the quality of the observations.

The lens B1608+656 was discovered using the Very Large Array
during a directed search for radio
lenses by the Cosmic Lens All-Sky Survey (CLASS) \citep{mye95}; two months later
it was rediscovered
in another independent survey aimed at studying faint peaked radio sources 
\citep{sne95}.
The source was found to consist of four flat-spectrum components
arranged in a typical ``quad'' configuration, 
with a maximum separation of 2.1'' between components.
Follow-up optical and infrared observations showed a similar morphology
and revealed the lensing galaxy, confirming 
the lens hypothesis \citep{mye95,fas96}.
Optical spectroscopy with the Palomar 5~m telescope 
measured a redshift of z$_l$=0.6304 for the lensing galaxy \citep{mye95},
and gave a redshift of  
z$_s$=1.394 for the source, which was also identified as a post-starburst
or E+A galaxy \citep{fas96}. Monitoring of the radio 
variability of the source allowed the  
measurement of the three independent time
delays between the components \citep{fas99}.

Several models reproducing the image positions, relative fluxes and
relative time delays in B1608+656
have already been constructed \citep{mye95,bla96,koo99,sur00}.
One conclusion obtained from these attempts is that 
modeling of the point source properties
is under-constrained and several solutions are possible \citep{sur00}. 
Additional constraints, from
the extended emission of the source, need to be incorporated
to break this degeneracy and allow for an accurate lens model 
and determination of the Hubble constant 
from the measured time delays.

High resolution optical and near infrared imaging of B1608+656 
with the Hubble Space Telescope (HST) has been acquired 
through four different filters.
In this paper we present an analysis of the 
V, I, and H-band archive exposures of B1608+656 
from HST proposals 6555 and 7422. 
This set offers a good multi-wavelength image sample
of B1608+656 in terms of signal-to-noise. 
The high resolution obtained in the three bands
reveals the extended structure of B1608+656.
We compare the images and present a diagnostic of 
wavelength-dependent distortions in the surface brightness of the
source, that are superposed on the distortions generated by the
gravitational lens deflections. 
Once identified, the chromatic effects need to be corrected
before using the extended emission of the source as a constraint
on lens model. 
The reconstruction of the images is the subject of
a coming paper 
(G. Surpi \& R. Blandford, in preparation, hereafter Paper 2).

The organization of this paper is as follows. 
In \S 2 we discuss the B1608+656 observations, starting with 
a summary of the results from radio monitoring in \S 2.1, 
followed in \S 2.2 by
a report of the optical and infrared exposures already
taken with HST.
\S 3 describes the process applied to
combine the F606W, F814W and F160W exposures, 
and compares the V, I, and H band images obtained. 
In \S 4, we estimate the age of the post-starburst population in the source
and adopt a model for its spectrum.
In \S 5,
we deconvolve the images using Tiny Tim generated point-spread-functions (PSF)
and the Lucy-Richardson method for deconvolution. 
\S 6 describes V$-$I and I$-$H color maps and the evidence
of extinction in the system. 
In \S 7, we discuss 
the properties of the optical and infrared images 
of B1608+656 and their potential use in the modeling 
of the lens mass distribution.
Appendix A contains details on the processing of the V and I images, and Appendix
B on the processing of H image.

\section{OBSERVATIONS OF B1608+656}

\subsection{Radio Properties}
\label{secradio}

The gravitational lens B1608+656 was discovered 
with the VLA at 8.4 and 15 GHz by two independent radio surveys  
in 1994 \citep{mye95,sne95}.
The radio images show the system consists of 
four well-separated components A, B, C, and D, 
all of them having flat radio spectra.
The radio positions and flux densities of the components, 
as determined by recent VLA 8.4 GHz observations of the system \citep{fas99}, 
are listed in Tab.~\ref{posfluxtd}. 
When a core of 51 mJy is subtracted at 1.4 GHz, 
the source is a radio galaxy 
with double-lobed structure having an overall size of $\sim$50''
and an integrated flux density of 12 mJy \citep{sne95}. 
The spectrum of the core is flat, while the double-lobe 
contribution can be modeled with a spectral index of 
$\alpha=0.8$ (${\rm F\propto \nu^{-\alpha}}$).
One of the lobes is highly polarized,
while the other one is unpolarized.
The core exhibited variability by up to 15\%
at a frequency of 8.4 GHz, allowing for the possibility to determine
the time delays in the system. 

%%%%%%%%%%%%%%%%%%%%%%%%%%%%%%%%%%%%%%%%%%%%%%%%%%%%%%%%%%%%
\begin{deluxetable}{ccccc}
\tablecolumns{5}
\tablewidth{0pc}
\tablecaption{Radio Properties of B1608+656 \label{posfluxtd}}
\tablehead{
\colhead{Component} & \colhead{Position ('')\tablenotemark{a}} & 
\colhead{Flux Density (mJy)\tablenotemark{b}} & 
\colhead{Relative Flux\tablenotemark{b}} & 
\colhead{Time Delay (days)\tablenotemark{b}}}
\startdata
A & (0.0000, 0.0000) & 34.29 & 2.042  & 31 $\pm$ 7  \\
B & (-0.7380, -1.9612) & 16.79 & 1.000  & 0           \\
C & (-0.7446, -0.4537) & 17.41 & 1.037  & 36 $\pm$ 7  \\
D & (1.1284, -1.2565) &  5.88 & 0.351  & 76 $^{+9}_{-10}$ \\ 
\enddata
\tablenotetext{a}{Position offsets with respect to component A in Cartesian 
coordinates, positive $x$-axis points west \citep{fas99}.}
\tablenotetext{b}{Results from the first season of VLA monitoring \citep{fas99}.}
\end{deluxetable} 
%%%%%%%%%%%%%%%%%%%%%%%%%%%%%%%%%%%%%%%%%%%%%%%%%%%%%%%%%%%%

A VLA 8.4 GHz monitoring program to determine the time delays of B1608+656
started in 1996. During the first season, 
extending from 1996 October to 1997 May, 64 observations 
were taken at 8.4 GHz \citep{fas99}. 
The light curve of the components showed flux-density variations 
at the 5\% level with common features in all four light curves,
including a rise of 5\% in the flux density, 
followed by a 20-day plateau and a drop of 4\%.
Even though the variations were small compared to those seen
in other lens systems, they allowed for a determination of
the three independent time delays 
at 95\% confidence level \citep{fas99},
as listed in Tab.~\ref{posfluxtd}.
During the second season of VLA monitoring  at 8GHz between February and October 1998
the components experienced a nearly
monotonic decrease in flux, on the order of 40\% \citep{fas00}.
The third season of monitoring has also now been completed.
The analysis of these new data 
should reduce the uncertainties on the time delays obtained in the first season.

\subsection{Summary of Optical and Infrared Imaging}
\label{secoptinf}

B1608+656 has been observed in the optical and infrared 
with the Hubble Space Telescope in three bands:
V-band (filters F555W and F606W), I-band (filter F814W) 
and H-band (filter F160W)
as summarized in Tab.~\ref{hstobserv}.
In this section we compare
the sets of observations at each band, and identify 
the best set of exposures in terms of signal-to-noise
and resolution to perform a multi-wavelength analysis of the lens.
 
The first set of frames in V and I bands was taken 
by Jackson 
in 1996 April, 
with the Wide Field Planetary Camera 2 instrument (WFPC2), through
the filters F555W and F814W (HST proposal 5908).
These images are presented in \citet{jac97}. 
A second set of exposures in V and I bands was obtained 
by Schechter
in 1997 November using the WFPC2, through filters F606W and F814W
(HST proposal 6555). 
Images in both sets 
suffer from contamination with cosmic rays and,
when the cosmic rays are removed,
display noise in the lens region which is
dominated by the Poisson fluctuations in the number of counts.
In this regime, after combining n frames,
the signal-to-noise of the resulting image approximately grows
as $\propto {\rm (n\times t)^{1/2}}$, 
where t is the exposure time of the frames
(see Appendix A). 
Due to their longer exposure time and the larger number of frames, 
we find the V and I-band images 
in HST proposal 6555
have S/N higher than those 
in HST proposal 5908
by factors of 3.5 and 2.5 respectively.

B1608+656 has also been observed in the infrared H-band 
with the Near Infrared Camera and 
Multi-Object Spectrometer (NICMOS) 
instrument of the HST through the filter F160W. 
The first observations were taken with NICMOS camera 2 in 1997 September  
by Falco, for the Cfa-Arizona-(H)ST-LEns-Survey (HST proposal 7495), 
see \citet{CASTLES}\footnotemark.
\footnotetext{CASTLES Gravitational Lens Data Base is available at: 
http://cfa-www.hardvard.edu/glensdata/B1608.html}
A second set of observations was conducted by Fassnacht in 1998 February,
as part of HST proposal 7422 (Readhead PI) using NICMOS camera 1.
Comparing both sets, due to the different pixel size in cameras 1 and 2, the
NIC1 frames have resolution higher by a factor 1.7.
On the other hand, NIC2 frames were acquired with dithering 
that can recover spatial resolution, but sometimes at the expenses 
of creating spatially correlated noise.
NIC1 frames have higher background noise compared to NIC2 frames
due to contamination of the lens with the high noise region of 
that camera. However, the total exposure time in NIC1 frames is longer
by a factor 8, and this increases the 
signal-to-noise of the combined NIC1 image.

We select the exposures from proposals 6555 and 7422
to perform our multi-wavelength analysis in V, I, and H bands,
for the following reasons: 
(i) In V band, the two sets of exposures in filters F555W and 
F606W are too close in wavelength 
to provide independent information about extinction, 
so we take the one with larger S/N.
(ii) In I band the combination of the 
two sets of exposures through filter F814W is not convenient.
Due to the relative rotation between the sets,
it would be hard to create
a combined PSF to deconvolve the combined image.
In addition, the information provided by the frames 
in proposal 5908 would not compensate the error introduced to register them.
(iii) In H band, NIC1 and NIC2 sets can probably give
images of comparable resolution and signal-to-noise.
Their combination is again not convenient for future deconvolution,
and we use the frames in NIC1 here.
In summary, the observations that we do not use in this study
are entirely consistent with those that we do use but do not 
improve the signal-to-noise.

%%%%%%%%%%%%%%%%%%%%%%%%%%%%%%%%%%%%%%%%%%%%%%%%%%%%%%%%%%%%
\begin{deluxetable}{lcccccc}
\tablecolumns{7}
\tablewidth{0pc}
\tablecaption{Optical and Infrared HST Observations of B1608+656 \label{hstobserv}}
\tablehead{
\colhead{Proposal PI} & \colhead{Proposal ID}& \colhead{Date} & 
\colhead{Instrument} & \colhead{Filter} & 
\colhead{Exposures} & 
\colhead{ExpTime(sec)}}
\startdata
N. Jackson & 5908 & 1996 Apr 7 & WFPC2  & F555W & 1 & 2 \\
&        &      &        &       & 3 & 500 \\
\cline{5-7}\\
&        &      &        & F814W & 3 & 800 \\
\cline{1-7}\\
E. Falco   & 7495 & 1997 Sep 29 & NIC2 & F160W & 4 & 704 \\
\cline{1-7}\\
P. Schechter & 6555 & 1997 Nov 1  & WFPC2 & F606W & 4 & 2900 \\
\cline{5-7}\\
&             &      &       & F814W & 1 & 2800 \\
&             &      &       &       & 3 & 2900 \\
\cline{1-7}\\
A. Readhead & 7422 & 1998 Feb 7 & NIC1 & F160W & 5 & 3840 \\
&                            &  & & & 1 & 2048 \\
&                            &  & & & 1 & 896 \\
\enddata
\end{deluxetable} 
%%%%%%%%%%%%%%%%%%%%%%%%%%%%%%%%%%%%%%%%%%%%%%%%%%%%%%%%%%%%

\section{PRELIMINARY ANALYSIS OF V, I, AND H-BAND IMAGES}
\label{secvih}

This section presents the results from archival HST data 
of B1608+656 from proposals 6555 and 7422.
Exposures include four frames obtained with the WFPC2 through filter F606W (V band),
four frames obtained with the WFPC2 through filter F814W (I band), and seven frames
obtained with NIC1 through filter F160W (H band). The exposure times
are listed in Tab.~\ref{hstobserv}.

Our goal is to identify qualitative
differences between the images and understand their origin.
Differences which are not intrinsic to the source but
due to external wavelength dependent processes, such as 
extinction or PSF convolution, have to be corrected 
before incorporating the extended structure of the source 
and lens galaxies as constraints on the modeling of the system. 
We want to compare the emission in V, I, and H bands 
on a pixel by pixel basis. 
Since we are interested in pixel photometry, and not just
aperture photometry, we keep the resolution 
of the images as high as possible.
Below we enumerate the steps followed to process and 
combine the frames at each band with this premise. We refer the reader 
to the Appendices A and B for more technical details.

We first estimate the mean sky level at each frame and subtract it.
In H frames a sky gradient, created by the propagation of the 
high noise region of NIC1 in the lens area, is also subtracted. 
We then proceed to register the frames.  
Only one geometrical transformation is performed in each frame.
The V and I band frames are registered 
by means of shifts. The H frames have smaller pixel size and are 
rotated with respect to V and I frames, so a general geometrical 
transformation is applied to them.  
We point out that only the centroids of components A and B 
are used to register the frames in different bands because,
as we will see in \S 6, 
the centroids of C, D, G1 and G2 can be apparently displaced
due to extinction. 
Finally we use averages to combine the frames at each band after masking
bad pixels and cosmic rays. 

\begin{figure}
	\figurenum{1}
       \epsscale{0.82}
%        \plotone{/home/surpi/plot/plotv.ps}
        \plotone{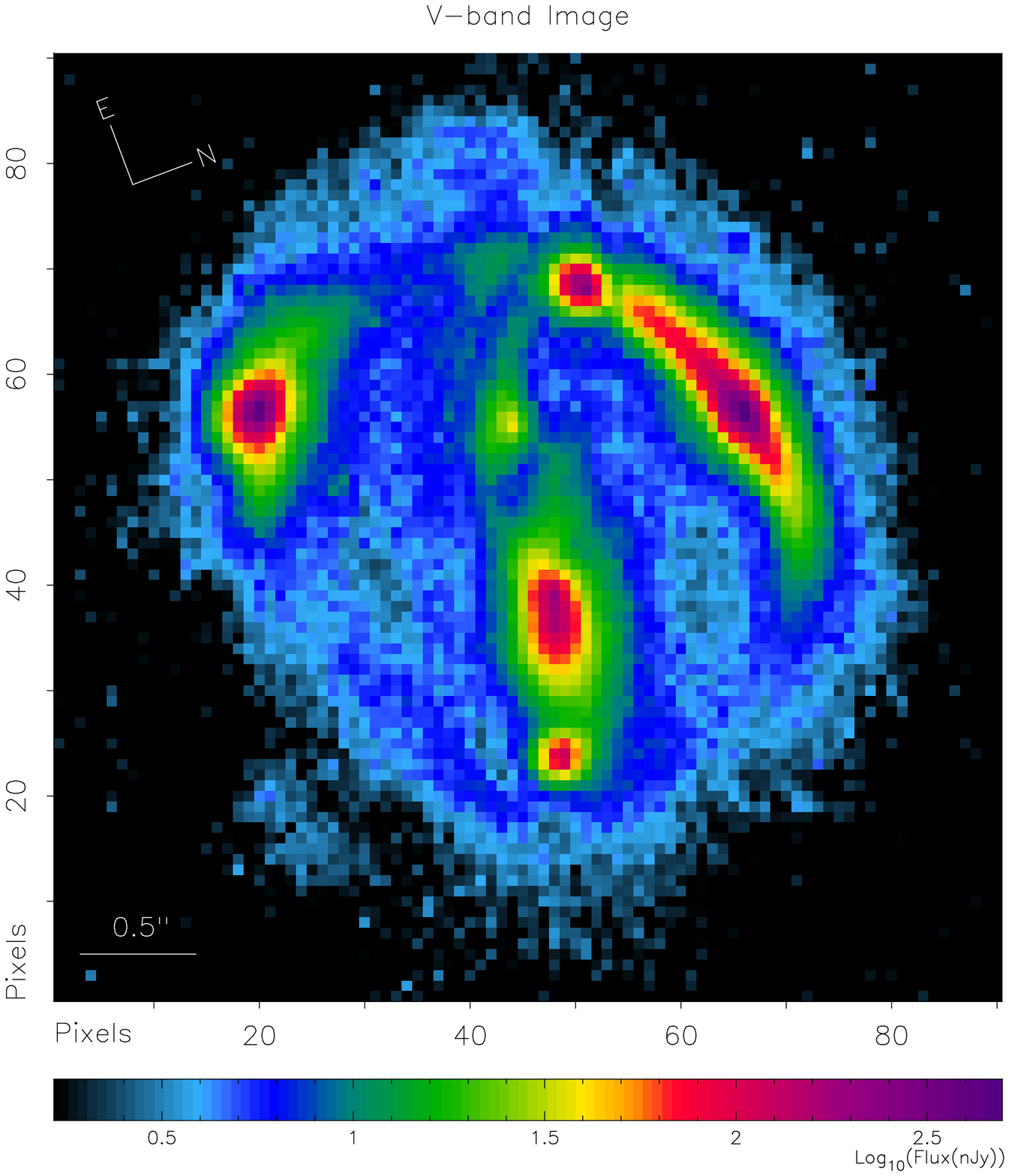}
       \caption{V band image of B1608+656 from F606W exposures by 
Paul Schechter (HST proposal 6555). Image is the average combination of 4 frames with a total
exposure time of 3 hrs 13 min. Only fluxes above 3 sigma of 
the sky noise are displayed.} 
       \label{plotv}
\end{figure}

\begin{figure}
	\figurenum{2}
       \epsscale{0.82}
%        \plotone{/home/surpi/plot/ploti.ps}
        \plotone{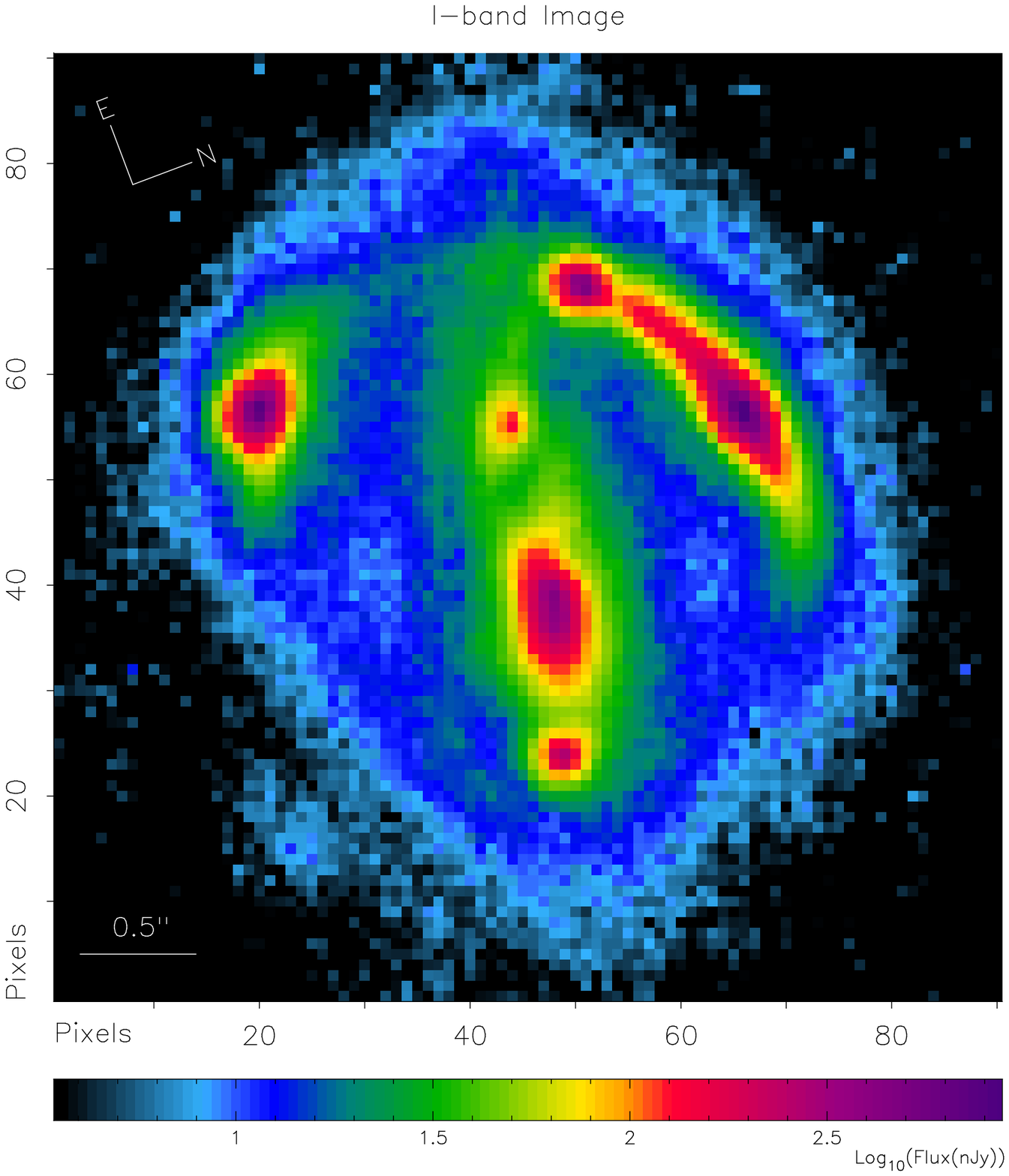}
       \caption{I band image of B1608+656 from F814W exposures by 
Paul Schechter (HST proposal 6555). Image is the average combination of 4 frames with a total
exposure time of 3 hrs 12 min. Only fluxes above 3 sigma of 
the sky noise are displayed.}
       \label{ploti}
\end{figure}

The final V, I, and H band images
are shown in Figures \ref{plotv}, \ref{ploti} and \ref{ploth}.
The orientation of the images is the original orientation of the V and I frames
(notice that north is not up).  
The intensity is in logarithmic units of nJy per pixel, 
and only fluxes above 3 $\sigma_{\rm s}$ are displayed, where 
$\sigma_{\rm s}=1.15, 0.55$ and 6 nJy are the mean standard deviation 
of the sky noise in V, I, and H bands respectively.
The H band image is the most affected by PSF convolution,
as shown by the clearly defined diffraction rings
surrounding A, C, and D components in Fig.~\ref{ploth}.
With a source redshift $z_s=1.394$, 
the filters F606W (V band), F814W (I band) and F160W (H band)
correspond to source emission at mean rest-frame wavelengths of 2506\AA, 3340\AA~
and 6713\AA, respectively. Similarly, 
for the lens galaxies located at $z_l=0.63$, we are observing the emission from
3882\AA, 4905\AA~ and 9860\AA~ in their rest frame.
The centroids of the four components of the source A, B, C, and D, 
and the lensing galaxies G1 and G2
in each band are listed in Tab.~\ref{tabcent}.

We evaluate the signal-to-noise in the images on a pixel-by-pixel basis.
In the V and I images the noise is estimated from contributions of 
Poisson, readout and background noise.
In H band the background noise changes from frame to frame,
and also across each frame, so we determine the noise
using the dispersion of the data at each pixel.
Contours of constant signal-to-noise for the images are plotted
in Fig.~\ref{plotstn}. In general V and I band have approximately 
the same level of S/N.
Compared to H, the S/N in V and I is higher in low emission regions. 
For example, low emission at 3 $\sigma_{\rm s}$ 
level has S/N$\sim$10, 7 and 3 in V, I, and H images respectively.
Isophotes enclosing 50\% of the source flux have approximately S/N$\sim$ 30 
in the three bands, and isophotes  
enclosing 10\% of the source flux have S/N$\sim$ 70, 70, 150 in the 
V, I, and H band respectively.

All bands reveals the extended structure of B1608+656,
showing the four multiple images of the background 
source embedded in a ring-like emission surrounding 
the two lens galaxies.
There are qualitative 
differences between the extended emission of the images, however.
First the saddle point 
that indicates the boundary between C and A images is closer to image C 
in V and I-band than in H-band. 
Since the separation between A and C is determined
by the critical curve associated with the lens potential, 
it should not depend on the wavelength.
The shift of 0.16'' in the observed saddle point can be a consequence of extinction,
PSF convolution, or a combination of both. 
If the shift is mainly due to high extinction near C,
then the actual saddle point would lie closer to H location.
On the other hand, if the shift is due to the prominent PSF Airy ring in H band
around image C,
then the actual saddle point would lie closer to its location in V and I band.
When the extended 
structure of the source is incorporated into the modeling 
of the system \citep{bla00}, the 
saddle points between multiple images (or flux minima in the Einstein ring) 
can place strong constraints on the lens potential.    
However, until the images are corrected for extinction and PSF distortions, 
the uncertainty in the location of the saddle point mentioned above 
can possibly lead to 
%misleading 
wrong conclusions if used to constrain or 
confront lens models as attempted by Kochanek, Keeton \& McLeod (2001).

The second qualitative difference between the images is 
that the centroid of the lens galaxy G1 appears 
to shift at decreasing wavelength in, approximately, the west direction.
The G1 centroid is displaced 
by 0.085'' in V band, and by 0.073'' 
in I band, with respect to its H band location
(see Tab.~\ref{tabcent}). There are two possible 
explanations. One is that the east part of the galaxy is being reddened,
in which case the G1 centroid would lie closer to its H band position. 
The other explanation, proposed by \citet{koo99},
is that the shift is a consequence of a change
in the intrinsic color of G1. 
These authors propose that dynamical interaction of
G1 and G2 could create a bluer region of star formation close to the
centroid of G1 in V band, and that this centroid 
would then represent the center of mass of the lensing galaxy. 
This interpretation supports their best fitting lens model, 
which was obtained using the centroid of G1 in V band as a constraint. 
However, until we understand the origin of the shift of the G1 center
as a function of the wavelength,
its use to constrain lens models can lead to erroneous conclusions.
We should notice at this point that the centroid of G1 is not alone in changing from one band to another; the centroids of G2, D,
 and C also 
shift as can be seen from Tab.~\ref{tabcent}. 

The main conclusion is that a correct interpretation
of the features observed in the extended structure of the background source
and lens galaxies at V, I, and H bands
requires further study of the PSF convolution, 
intrinsic color variation and extinction in the system. 
We discuss these effects in the next sections.

%%%%%%%%%%%%%%%%%%%%%%%%%%%%%%%%%%%%%%%%%%%%%%%%%%%%%%%%%%%%
\begin{deluxetable}{cccc}
\tablecolumns{4}
\tablewidth{0pc}
\tablecaption{Centroids\label{tabcent}\tablenotemark{~}}
\tablehead{
\colhead{} &\colhead{V-band} &\colhead{I-band} &\colhead{H-band}
} 
\startdata
A &  65.74$\pm$0.05~~~ 56.62$\pm$0.06 & 65.84$\pm$0.04~~~ 56.55$\pm$0.05 & 65.64$\pm$0.03~~~ 56.51$\pm$0.03 \\ 
B &  19.88$\pm$0.05~~~ 56.57$\pm$0.06 & 19.94$\pm$0.04~~~ 56.60$\pm$0.05 & 19.92$\pm$0.03~~~ 56.62$\pm$0.03 \\
C &  50.57$\pm$0.07~~~ 68.50$\pm$0.07 & 50.67$\pm$0.05~~~ 68.40$\pm$0.05 & 50.92$\pm$0.04~~~ 68.31$\pm$0.03 \\
D &  48.72$\pm$0.10~~~ 23.81$\pm$0.10 & 48.79$\pm$0.07~~~ 23.83$\pm$0.07 & 48.32$\pm$0.04~~~ 23.63$\pm$0.04 \\
G1 & 48.02$\pm$0.06~~~ 37.55$\pm$0.13 & 47.98$\pm$0.04~~~ 37.83$\pm$0.09 & 47.18$\pm$0.04~~~ 39.22$\pm$0.05 \\
G2 & 43.50$\pm$0.19~~~ 55.47$\pm$0.27 & 43.62$\pm$0.10~~~ 55.33$\pm$0.14 & 44.13$\pm$0.06~~~ 55.26$\pm$0.07 \\ 
\enddata
\tablenotetext{~}{At each band the two values correspond to
$\left(x,y \right)$ positions measured in pixel coordinates from images
in Figs. \ref{plotv}, \ref{ploti}, and \ref{ploth}, 1~pixel=0.0455''.
Centroids and errors were determined
using the IMCENTROID routine \citep{sts01}.}
\end{deluxetable} 
%%%%%%%%%%%%%%%%%%%%%%%%%%%%%%%%%%%%%%%%%%%%%%%%%%%%%%%%%%%%

\section{SOURCE POST-STARBURST POPULATION}
\label{post-starburst}

\begin{figure}
	\figurenum{5}
       \epsscale{0.6}
%        \plotone{/home/surpi/plot/plotspecB.ps}
        \plotone{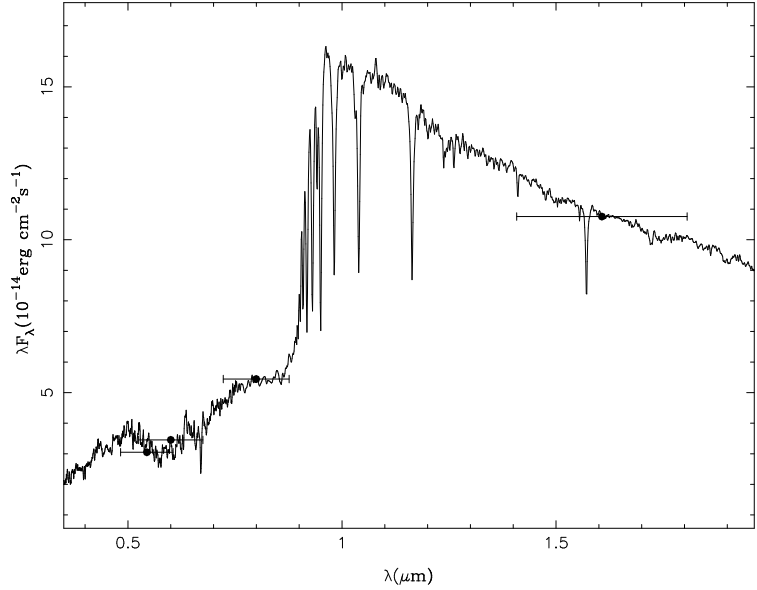}
       \caption{Photometry of the B image of the source through
filters F555W, F606W, F814W (HST, WFPC2, PC) and F160W (HST, NIC1).
Horizontal error bars show the effective width of the WFPC2
filters, and the FWHM of the NIC1 filter.
The line shows the 500 Myr old population model from 
\citet{bru93} redshifted to the observer frame. 
}
       \label{specB}
\end{figure}

\begin{figure}
	\figurenum{6}
       \epsscale{0.8}
%        \plotone{/home/surpi/plot/plotabsorB.ps}
        \plotone{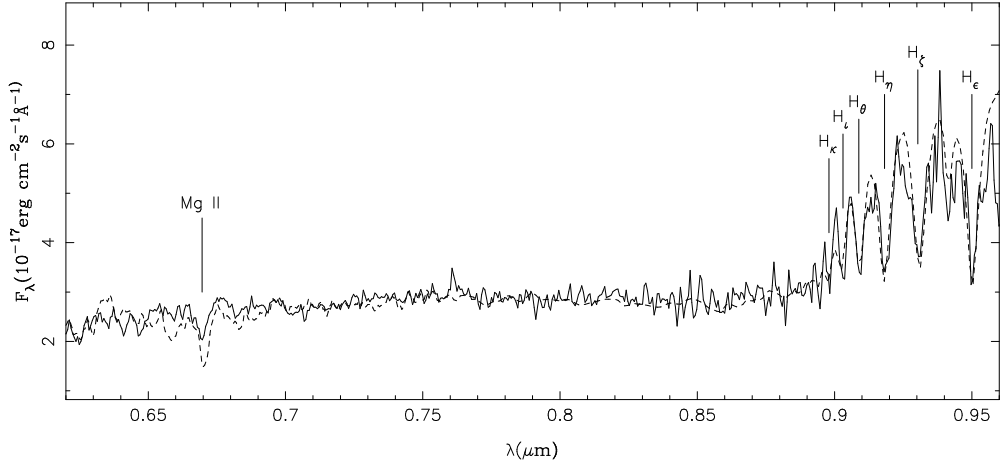}
       \caption{Comparison of the optical spectrum of the source in
the gravitational 
lens B1608+656 measured by \citet{fas96} (solid line)
with the 500 Myr old population model of \citet{bru93}
(dashed line). Wavelengths are as observed in air.
The vertical lines mark the position of the absorption features 
of the source detected by \citet{fas96}, which locates it
at $z=1.394$.
}
       \label{absorB}
\end{figure}

An optical spectrum of the background source in B1608+656 was obtained 
with the double spectrograph on the Palomar 5 m telescope
by \citet{fas96}, by 
positioning the slit so as to minimize the lens galaxies light. 
The spectra showed prominent high-order Balmer absorption lines 
and Mg II absorption,
allowing the determination of 
a conclusive redshift of $z=1.394$ for the source,
and indicating it is a post-starburst or E+A galaxy.
No emission associated with the AGN was found.
Comparing photometry of the source 
through the filters F555W, F606W, F814W and F160W
with isochrone synthesis models
of \citet{bru93} we find that the post-starburst population
in the host galaxy is around 500 Myr old. 
Fig.~\ref{specB} shows    
photometry of the B component of the source within an
elliptical aperture of radius 0.3'', superposed on the spectrum model.
There is a remarkable agreement of the photometry points with the model.
The same fit is found for photometry in different areas of B,
as this image shows no significant change of color. 
When the optical spectrum of the source in \citet{fas96} is compared with the model,
we find the absorption lines and overall shape of the continuum of the 
source are fitted well by the 500 Myr old instantaneous burst model of \citet{bru93} as showed in Fig.~\ref{absorB},
confirming the age estimation.
%The agreement of the photometry and spectra with the model confirms
%the relative absence of dust around image B.

This model can set an upper limit to the age of the starburst population
in the source. Older starbursts would be redder than the source, and are
ruled out since no mechanism would be able to account for the discrepancies.
On the other hand, younger starburst models, if reddened or combined with 
older stellar components, could still fit the source. 
We experimented with a younger starburst subject to extinction or 
superposed on older populations
but the fit to the spectrum and photometry of the source was inferior.
We adopt the 500 Myr old instantaneous starburst model to describe the 
stellar population at the source in what follows. The model provides
a synthetic spectrum for the source 
in the violet end of visible light and infrared where no direct 
observations are available.
The spectrum is used in the next section to generate the PSFs of 
the V, I, and H filters.

\section{IMAGE DECONVOLUTION}
\label{secdeconvolution}

In this section we present initial deconvolutions of the V, I, and H band images.
We use the Tiny Tim software package to generate HST model Point Spread Functions 
\citep{kri99}\footnotemark, 
\footnotetext{Tiny Tim User's Guide is available at: 
http://www.stsci.edu/software/tinytim}
and the Lucy routine in IRAF \citep{sts01}\footnotemark to deconvolve the images.
This task restores images using the Lucy-Richardson method \citep{luc74}
adapted for HST images contaminated with Poisson noise.
\footnotetext{Space Telescope Science Data Analysis System is available at: 
http://ra.stsci.edu/STSDAS.html}

HST PSFs comprise a core,
diffraction rings and radial spokes. Their shape is dominated by diffraction 
rather than seeing effects, and the PSFs appearance 
changes at increasing wavelength as the diffraction patterns expand proportionally. 
In broad-band filters, as F606W, F814W, and F160W, 
the superposition of the monochromatic Airy rings at the different wavelengths of the bandpass
results in PSFs with smooth wings. 
The generation of these polychromatic PSFs           
requires knowledge of both the transmission function of the filter and the 
spectrum of the observed object. 
We use the database 
of wavelength and weights supplied by the Tiny Tim package for the transmission
of each filter, and the 500 Myr old instantaneous starburst model
from \citet{bru93} to characterize B1608+656 spectrum.

PSFs are constructed with a resolution of 0.011'' by subpixelizing with a factor 4 
in each dimension. We subpixelize the images by the same factor using linear
interpolation and deconvolve them with the Lucy routine in IRAF \citep{sts01}$^5$.
The criterion adopted for convergence was to increase the number of iterations
until the deconvolved image convolved with the PSF 
reproduced the observed image within an error of 5\%.
Many light distributions are, after convolution with the PSF, compatible within a 5\%
error with the observed image. It is expected then that 
the difference between the
deconvolved image and the original light distribution is higher than 5\%.  
After deconvolution the images were repixelized to its original resolution.
The deconvolved images are shown in Fig.~\ref{plotdec}. 
The H band image is the most difficult one to deconvolve,
since the diffraction rings are very prominent. Residuals of the rings 
are still present around A, C, and D in the deconvolved H image.  
The H deconvolution did not solve for the uncertainty in the 
saddle point between A and C components. 
The arc in the deconvolved H image now shows two minima, 
one at each side of the intersection between the remnants of 
the Airy ring around C and the arc.
The two locations correspond approximately to the position 
of the minima in V and I, and 
the minima in H before deconvolution. 

We should stress here that deconvolution is not a unique operation, 
and the images presented here are the best results obtained 
under certain limitations: (i) 
The model assumed for the spectrum is known to fit the observed spectrum
of the source, which lies between 6000 \AA~ and 9600 \AA~.
If the extrapolation of the model toward lower and higher wavelengths
has local discrepancies with the spectrum of the source, the 
restoration in V and H bands might be limited.
(ii) The spectrum was assumed constant. It was modeled after the source 
to obtain better results in the deconvolution of the four images of that object.
However, as the result of contamination with light from the lens galaxies
and extinction, the color does change across the frame as we will show 
in the next section. 
(iii) Even for sources of known constant color, 
deconvolutions are limited by the accuracy of the algorithms used to 
deconvolve. We tried an alternative deconvolution method,  
the Maximum Entropy Method (MEM package in IRAF \citep{sts01}$^5$), and found similar
results, except that it produced a slightly higher noise level. 
The creation of artifacts by the amplification of the noise is the principal
illness of the standard methods for deconvolution. 

\section{EXTINCTION AND COLOR MAPS}
\label{extincsec}

The three HST images have different angular resolutions
because they were convolved with different PSFs during observation.
The deconvolution probably has restored the images to comparable resolution,
but has also amplified their noise.
Using the deconvolved images to create the color maps
results in maps which are dominated by pixel-scale structure.
However, the extinction is an average over a galactic scale length 
along the line of sight, so it ought to vary relatively slowly across the galaxies.
The small scale structure observed has to be artificial, due to noise that propagates
from the deconvolved images. To reduce the noise in the color maps,
instead of using deconvolved images, for our analysis we created
images of comparable resolution by further convolving each image.
At each band we create a PSF$\rm{_{ratio}=PSF_{Gauss}/PSF_{Filter}}$, 
given by the ratio (in Fourier space) between
a Gaussian PSF with $\sigma$=1.5 pixels, and the PSF of the corresponding filter. 
The convolution of the observed images with 
the ``PSF$_{\rm ratio}$'' 
results in images of comparable resolution, equivalent to the original images
convolved with the Gaussian PSF. In this section we will use
V, I, and H band images that have been subjected this procedure.

A preliminary inspection of the relative magnifications of components A, B, C, and D
at H, I and V bands reveals increasing extinction
as we move to shorter wavelengths.
We select the $N=90$ brightest pixels in image B at each band. 
These pixels are delimited by an isophote of ellipticity 0.34 
and semi-major axis of 0.3'',
and enclose fluxes of 6.90, 14.5 and 57.6 $\mu$Jy at V, I, and H band respectively
(which is approximately 50\% of the total flux in component B).
Since image B is the furthest away from G1 and G2,
we assume that it is the least reddened and 
we compute the extinction of A, C, and D relative to B.
Assuming the measured radio flux ratios (listed in Tab.~\ref{posfluxtd})
are correct we take the brightest 
$N$ pixels, with $N=90\times (2.042,1.037,0.35)=183.8, 93.3, 31.5$ for A, C, and D.
These pixels cover matching areas at the source
and their relative fluxes should reproduce, in absence of extinction, 
the observed radio magnifications. For example, at V band the pixels selected at A
should add to flux ${\rm F}_V(0)=2.042\times 6.90\mu$Jy$=14.1\mu$Jy.
We compute the relative extinction at each component 
as ${\rm A}_\lambda=-2.5 \log_{10}({\rm F}_\lambda/{\rm F}_\lambda(0))$, 
where $\lambda=$ H, I and V and ${\rm F}_\lambda$ is the 
flux measured in the selected patch.
The values are listed in the Tab.~\ref{extinc} and show that
A suffers little extinction, while C and D are strongly reddened relative to B.
 
%%%%%%%%%%%%%%%%%%%%%%%%%%%%%%%%%%%%%%%%%%%%%%%%%%%%%%%%%%%%
\begin{deluxetable}{ccccc}
\tablecolumns{5}
\tablewidth{0pc}
\tablecaption{Extinction of the Source\label{extinc}\tablenotemark{a}}
\tablehead{
\colhead{} &\colhead{A} &\colhead{B} &\colhead{C} &\colhead{D}} 
\startdata
${\rm A_H}$ & 0.08 & 0.0  & 0.14 & 0.20 \\
${\rm A_I}$ & 0.08 & 0.0  & 0.43 & 0.42 \\
${\rm A_V}$ & 0.12 & 0.0  & 0.69 & 0.71 \\
\enddata
\tablenotetext{a}{Mean extinction relative to B component. 
The A, C, and D extinctions are 
measured over areas that match elliptical aperture of 0.3'' in B.}
\end{deluxetable} 
%%%%%%%%%%%%%%%%%%%%%%%%%%%%%%%%%%%%%%%%%%%%%%%%%%%%%%%%%%%%
    
\begin{figure}
	\figurenum{8}
       \epsscale{1.0}
%       \plotone{/home/surpi/plot/plot2col.ps}
       \plotone{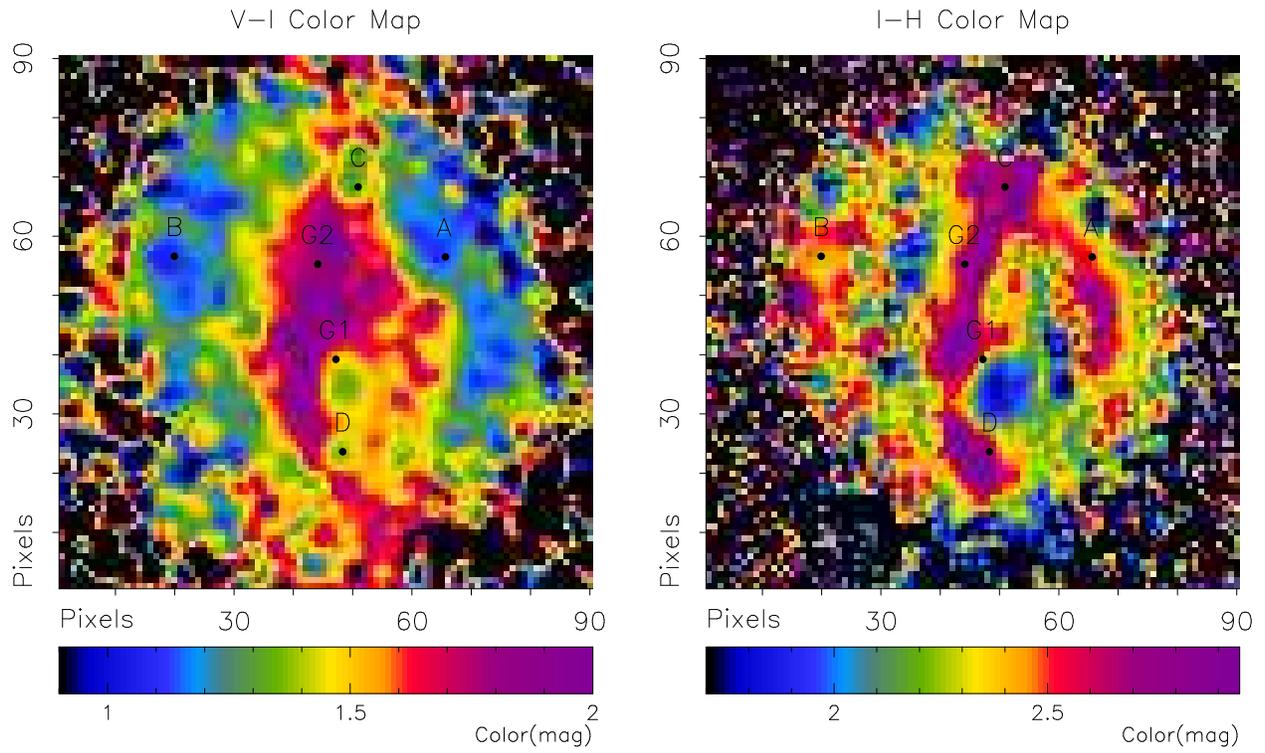}
       \caption{V$-$I and I$-$H color maps. 
V, I, and H band images have been convolved with 
the ratio between a Gaussian PSF and the PSF of the corresponding filter,
to achieve same resolution before combining them into the color maps.}
       \label{plotcol}
\end{figure}

There are two independent colors, which we choose to be V$-$I and I$-$H.
We use Vega: magnitudes m=$-2.5\log _{10}({\rm CR})+{\rm ZP}$, 
where CR is the count rate
(measured counts per second) and the zero point ZP=21.49, 21.64, 22.89
for F160W, F814W, F606W respectively 
\citep{sts98}. 
Colors are then evaluated pixel by pixel
as V$-$I$={\rm m_V-m_I=-2.5\log_{10}(V/I)+1.25}$ and 
   I$-$H$={\rm m_I-m_H=-2.5\log_{10}(I/H)+0.15}$, 
where images are in CR units.
V$-$I and I$-$H colors are shown in Fig.~\ref{plotcol}. 
The orientation of the color maps is as in the Figures \ref{plotv},
\ref{ploti} and \ref{ploth},
so that north is not up. To avoid confusion we will use up-down-left-right
to locate features in the maps. 

Color gradients in Fig.~\ref{plotcol}
can arise from two different effects, extinction or variation in the intrinsic 
color of the objects.
At first sight, there is an overall pattern in the color.
Both maps show maximum values
in a vertical stripe compromising C, G2, G1 and D. 
A change of color in C and D (with respect to A and B) was expected since both images are 
highly reddened. In G1 and G2 area, the stripe is broader in V-I than in I-H map,
showing the lenses have different intrinsic color than the source.
Contamination of C and D with light from the lens galaxies could also
be contributing to their change of color respect to A and B.
Estimated colors within an elliptical aperture of radius 0.3''
in B and matching areas in A, C, and D are reported in Tab.~\ref{tabcols}.
In V$-$I map the color shows larger gradients across the components
and minimum and maximum values are listed in Tab.~\ref{tabcols}, 
while I$-$H colors reflect average
values.

%%%%%%%%%%%%%%%%%%%%%%%%%%%%%%%%%%%%%%%%%%%%%%%%%%%%%%%%%%%%
\begin{deluxetable}{ccccc}
\tablecolumns{5}
\tablewidth{0pc}
\tablecaption{Source Color\label{tabcols}}
\tablehead{
\colhead{Color(mag)} & \colhead{A}&\colhead{B}& \colhead{C}&\colhead{D}}
\startdata
V$-$I\tablenotemark{a} & 1.10-1.50 & 1.06-1.28 & 1.28-1.60 & 1.38-1.51 \\
I$-$H\tablenotemark{b} & 2.52 & 2.47 & 2.70 & 2.48 \\
\enddata
\tablenotetext{a}{Colors are minimum and maximum values measured within an elliptical radial aperture of 0.3''
in B, and matching areas in A, C, and D.}
\tablenotetext{b}{Colors correspond to mean values within the same apertures.}
\end{deluxetable} 
%%%%%%%%%%%%%%%%%%%%%%%%%%%%%%%%%%%%%%%%%%%%%%%%%%%%%%%%%%%%

The maps in Fig.~\ref{plotcol} have color variation
across the lensing galaxy G1.
There is a minimum of color centered approximately at pixel 
(49,34), and an steep color gradient  
toward the upper-left direction. Minimum and maximum colors 
observed in G1 area are tabulated in Tab.~\ref{tabcolg}.
The bluer region of G1 does not correspond to the nucleus
of the galaxy, it is located 0.25'', 0.18'' and 0.17'' off the centroid of the
surface brightness distribution in H, I and V bands respectively 
(see Tab.~\ref{tabcent}).
But, if the two lensing galaxies were interacting dynamically 
creating a region of star formation in that area,
the change of color could still be interpreted 
as intrinsic to G1 as suggested by \citet{koo99}.
However, photometry through the bluer window shows that region is
not singular and follows a clearly defined 
de Vaucouleur profile \citep{bla00} identifying G1 as an elliptical galaxy.
The photometry also indicates that G1 is bluer
than the average
spectral energy distribution of normal ellipticals
(see for example \citep{sch97}). This would agree with the hypothesis of \citet{mye95}
that the lens might also be a post-starburst galaxy, 
based on the absorption lines observed in the spectrum.
We conclude that the color variations in G1 do not reflect 
a change of its intrinsic color but differential extinction.
The extinction is probably due to dust associated with G2,
since ellipticals usually contain little or no dust.
The reddest area in Fig.~\ref{plotcol}
shows the extinction by dust G2 has probably left behind 
when it swung around G1. 
%%%%% Roger comment:
%On the basis of these color maps we conclude that the dust
%now associated with G2 must extinguish more than half of the light 
%of G1, implying that galaxy G2 is closer to us than G1
%and has a smaller redshift.
%Although it is possible to contrive different models this is the most
%natural explanation.
On the basis of these color maps we conclude that the dust
now associated with G2 must extinguish more than half of the light 
of G1, implying that the dust is closer to us than G1.
Although it is possible to contrive different models the most
natural expectation is that galaxy G2 has a smaller redshift than G1.
Either way a measurement of $\sigma_{\rm G1}$ and
${\rm V_{G2}-V_{G1}}$ would be most instructive. 
%We show in Figure \ref{plotg2t} the possible trajectory   
%of G2 superposed to the I band image.

%The disposition of the reddest regions with 
%respect to the nucleus of G1 leave little doubt that G2 has swung by G1 and is 
%currently moving outward. 

The minimum colors quoted in Tab.~\ref{tabcolg} for G1 are seen through  
the lower reddening area and place an upper limit on G1's intrinsic color
(a constant reddening over G1 can still be taking place modifying its 
photometric color). Combining these values with the maximum colors reported we estimate
G1 color excess 
${\rm E(V-I)={\rm A}_V-{\rm A}_I=(V-I)_{\rm max}-(V-I)_{\rm min}=0.6}$, and
${\rm E(I-H)=1.1}$ in a similar way. 
G1 colors were also extracted by \citet{koc00}; 
they found ${\rm V-H=4.48\pm 0.23}$ and ${\rm I-H=2.18\pm 0.35}$
(not corrected for extinction), with V corresponding to F555W.
>From Tab.~\ref{tabcolg} we find mean values
${\rm (V-H)_{\rm mean}=3.97}$ and ${\rm (I-H)_{\rm mean}=2.32}$. 
The discrepancy between the V$-$H values is probably 
because the V band corresponds to different filters, 
F606W (here) and F555W (in citation).  
We don't expect the colors to agree perfectly for two further reasons.
First, the G1 color varies across the source and the average
used here might differ from the one used in \citet{koc00}.
Second we estimate colors after convolving the images with the corresponding
PSF$_{\rm ratio}$,
while in \citet{koc00} images have not been convolved.
We remark that mean colors are not representative of G1 intrinsic color,
nor of its reddened color, but just of its average.

In the case of G2, the color also varies across the galaxy.
G2's photometric centroid differs in each band, 
the centroid shifts approximately to the left direction
at decreasing wavelength. 
The I band centroid is shifted by 0.023'', 
and V centroid by 0.030'', from H centroid (see Tab.~\ref{tabcent}). 
This suggests that the extinction is higher in the right part
of the galaxy. The color maps in Fig.~\ref{plotcol} show a redder area
around pixel (46,56) which is 0.1'' to the right from G2 mean 
centroid in V, I, and H,
likely to be a region of maximum extinction in G2 rather than a 
redder intrinsic color. 
We then interpret the variation of color observed in G2 as mainly due 
to extinction,
and once again the minimum and maximum colors observed give indications of 
the intrinsic color
and the most reddened area respectively. G2 colors are reported 
in Tab.~\ref{tabcolg}, and from them we estimate
${\rm E(V-I)=0.3}$ and ${\rm E(I-H)=0.8}$. 
Since G2 is abundant in dust,
it is probably a late type galaxy, 
highly distorted due to its encounter with G1.

%%%%%%%%%%%%%%%%%%%%%%%%%%%%%%%%%%%%%%%%%%%%%%%%%%%%%%%%%%%%
\begin{deluxetable}{ccc}
\tablecolumns{3}
\tablewidth{0pc}
\tablecaption{Lens Galaxies Color\label{tabcolg}}
\tablehead{
\colhead{Color(mag)} &\colhead{G1} &\colhead{G2}}
\startdata
V$-$I\tablenotemark{a}& 1.34 - 1.97 & 1.67 - 2.00  \\
I$-$H\tablenotemark{a}& 1.77 - 2.87 & 2.10 - 2.94 \\
\enddata
\tablenotetext{a}{Colors are minimum and maximum values measured.}
\end{deluxetable} 
%%%%%%%%%%%%%%%%%%%%%%%%%%%%%%%%%%%%%%%%%%%%%%%%%%%%%%%%%%%%

The color maps showed in Fig.~\ref{plotcol} have two different sources 
of error. First, noise coming from the
V, I, and H band images (see Fig.~\ref{plotstn}). 
For the V$-$I map it can be estimated as 
${\rm (2.5/\ln{10})\left[ (S/N)_V^{-2}+(S/N)_I^{-2} \right]^{1/2}}$,
and similarly for the I$-$H map. This noise decreases at increasing signal
and it will mainly compromise the restoration of the low emission regions. 
The second uncertainty in the color maps comes from the two PSF convolutions, 
one during the observation and the second performed here. 
The noise pattern they produce is 
opposite to the previous one, with the noise increasing 
at increasing signal in the images.
The degradation due to the convolutions is 
significant and the only way of dealing with it is to 
apply the color maps to images with the same resolution, 
i.e. images that have been treated with the same convolution process.

\section{DISCUSSION}
\label{discuss}     

The V, I, and H band HST images of B1608+656 show 
Einstein ring emission from the four images of the radio source, 
encircling the two lens galaxies.            
Pixel photometry indicates qualitative differences between the emission
in the three bands, arising from extinction and PSF convolution.
It is immediately apparent from V-I and I-H color maps that 
most of the reddening is due to dust associated with the lens galaxy G2, 
probably a late type galaxy, which has been gravitationally 
swung around by the other lens galaxy, G1, an elliptical. 
The extinction 
most strongly affects G2, the east portion of G1,
and two of the four multiple images of the source, the ones
nearly aligned with the position angle of the lenses.
Of the three bands, H is the most affected by PSF convolution.
The H band image shows prominent Airy rings around the multiple images,
and is the most difficult to restore. 
The H band image in NIC2, from CASTLES, suffers from the same
problem. Additional infrared imaging with a 
large ground based telescope equipped with adaptive optics would
be worth pursuing. 

There are multiple features that can be seen in 
the extended emission and used as constraints 
to break the degeneracy of current lens models:
(i) the location and radial profile of the lens galaxies
(which are not observed in the radio), (ii) special properties of the 
Einstein ring like saddle points, traces of the critical curve,
and the inner and outer limits where quad images are formed, 
and (iii) the overall ring surface 
photometry, which satisfies a four-to-one mapping into the source plane.
However, the strong extinction in V and I bands, 
the significant PSF distortion in H band, and the contamination
of the ring with emission from the lens galaxies make clear that the images
need to be corrected before being used as constrains,
or they could lead to misleading conclusions. The 
multi-wavelength strategy required to correct the images
will be applied in Paper2, but it 
can be schematized here based on the observations we made above.
First, images have to be taken to comparable spatial resolution.
The extinction can then be measured from the color maps constructed with them.
If current methods for deconvolution are applied, 
as these techniques amplify the noise in the images,
the deconvolution should be done after the images have been corrected for extinction. 
PSF deconvolution and the measurement of extinction require
knowledge of the intrinsic color of the source, and yield better
results if applied to a source of constant color, so that
a decomposition of the surface photometry into emission from the lens galaxies
and the source host galaxy, should precede them. 

\acknowledgements  
We thank Paul Schechter and Chris Fassnacht for the acquisition of
the B1608+656 HST data analysed here
and for their encouragement and comments on this manuscript. 
We thank Eric Agol and Leon Koopmans for discussions and 
also comments on the presentation that led to an improvement 
in the manuscript. 
We are also grateful to Gustavo Bruzual for 
providing data of the starburst models, and Chris Fassnacht
for providing B1608+656 spectral data. This material is based upon work supported by the
NSF under award numbers AST-9529170 and  AST-9900866 and the NASA under contract 
number NAG5-7007.

\appendix 
\section{APPENDIX: V AND I-BAND IMAGE PROCESSING}
\label{appAsec}     

We process here the set of HST archival exposures 
from proposal 6555, taken in 1997 November  
using the planetary camera PC1 of the WFPC2 instrument, 
through the filters F606W and F814W (see Tab.~\ref{hstobserv}).
The PC1 has a pixel size of 0.0455'' 
and a format of 800x800 pixels. 
Pixel data are converted from DN counts units into
electrons multiplying by the factor of Gain=7e$^{-}$DN$^{-1}$. 
Each e$^{-}$ corresponds to one photon detected.
Beyond the quantum efficiency,
the efficiency of each WFPC2 exposure is limited by the photon 
count noise $\sigma_p$, the instrument read-out noise $\sigma_r$, 
and the sky background noise $\sigma_b$. The standard deviation
at each pixel is then $\sigma=(\sigma_p^2+\sigma_r^2+\sigma_b^2)^{1/2}$. 
For the WFPC2 the read-out noise standard deviation is $\sigma_r=5{\rm e}^{-}$. 
Photon counts result from a Poisson process in which the
standard deviation $\sigma_p$ can be estimated as the square root of the signal 
in photon counts S.
The signal-to-noise of the image then equals $S/N=S/(S+\sigma_r^2+\sigma_b^2)^{1/2}$.  
The S/N will approximately grow linearly with the signal 
strength for $S<\sigma_r^2+\sigma_b^2$, and
then as the square-root of the signal, as the Poisson noise in the object 
begins to dominate over the sky and read-out noise. Notice that in the last regime
the S/N increases as the square root of the exposure time. 

There are eight frames in proposal 6555,
hereafter designated as f1,f2... f8 according
to the time order they were observed.
Frames f1, f2, f5 and f6 were taken using the F814W filter 
and frames f3, f4, f7 and f8 with the F606W filter.
We measure and subtract the mean sky level at each frame.
The values obtained are listed in Tab.~\ref{vidata}.
V frames have a mean sky level of $48.7\pm 11.8$ photons.
The standard deviation of $\sigma=11.8$ photons
results from the combination of  
$\sigma_p=(48.7)^{1/2}=7.0$ photons, $\sigma_r=5$ photons, and $\sigma_b=8.1$ photons.
We find the I frames have a mean sky background of $26.3\pm 10.0$ photons
which yields $\sigma_b=7.0$ photons.

%%%%%%%%%%%%%%%%%%%%%%%%%%%%%%%%%%%%%%%%%%%%%%%%%%%%%%%%%%%%
\begin{deluxetable}{ccccc}
\tablecolumns{5}
\tablewidth{0pc}
\tablecaption{Sky Level and Shift in V and I-band Frames \label{vidata}}
\tablehead{
\colhead{Frame} &\colhead{Band}& \colhead{Sky Level\tablenotemark{a}} & 
\colhead{Multi-pixel Shift\tablenotemark{b}} & 
\colhead{Sub-pixel Shift\tablenotemark{b}}}
\startdata
f1 & I & 23.3 $\pm$ 9.9  & +5~~+5 & +0.05~~ -0.21  \\
f2 & I & 26.6 $\pm$ 10.0 & +5~~+5 & +0.08~~ -0.24 \\
f3 & V & 49.4 $\pm$ 11.8 & +5~~+5 & -0.13~~ +0.21\\
f4 & V & 49.0 $\pm$ 11.8 & +5~~+5 & -0.10~~ +0.20 \\
f5 & I & 27.9 $\pm$ 10.1 & +0~~+0 & +0.10~~ -0.20  \\
f6 & I & 26.5 $\pm$ 10.1 & +0~~+0 & +0.11~~ -0.20 \\
f7 & V & 47.8 $\pm$ 11.8 & +0~~+0 & -0.06~~ +0.21 \\
f8 & V & 48.6 $\pm$ 11.7 & +0~~+0 & -0.06~~ +0.22 \\
\enddata
\tablenotetext{a}{Sky measured in units of photons.} 
\tablenotetext{b}{The two values correspond to x and y shifts 
in the pixel grid.}
\end{deluxetable} 
%%%%%%%%%%%%%%%%%%%%%%%%%%%%%%%%%%%%%%%%%%%%%%%%%%%%%%%%%%%%

All the input frames in F606W and F814W are registered together before 
performing the average in each filter. 
We first perform the multi-pixel shifts listed in Tab.~\ref{vidata}. 
Prior to determine the sub-pixel shifts,
bad pixels and cosmic rays are replaced with median values at each frame.
We use component B 
in B1608+656 and a star located at $(2.96'' , 7.93'')$ from B
as references to determine sub-pixel shifts relative to f1
by cross-correlating the frames. Since the error introduced in a shift 
increases with the amount shifted,
we re-normalize our shifts to the minimum possible values. This also
keeps the resolution in the frames comparable. The (renormalized)
sub-pixel shifts applied are listed in Tab.~\ref{vidata}.

The frames are contaminated with bad pixels and cosmic rays.
To clean them before performing the average
we create a pixel mask for each frame 
rejecting bad pixels and pixels above and below 3 $\sigma$ level of the signal.
For the thresholding we estimate the signal S with the median, 
and $\sigma=(${\rm S}$+\sigma_r^2+\sigma_b^2)^{1/2}$, with the $\sigma_b$
above determined. 
After averaging the frames with weights according to their
exposure time, we arrive at the
final V and I images shown in Figs.~\ref{plotv} and \ref{ploti}.
We create signal-to-noise maps evaluating 
${\rm S/N=\sqrt{n}(S/\sigma)}$ on a pixel-by-pixel basis, where n
is the number of frames contributing to the average at each pixel
after masking.  Contours of constant S/N=10, 15 and 30 in V, 
and  S/N=7, 15 and 30 in I
are plotted in Fig.~\ref{plotstn}.

\section{APPENDIX: H-BAND IMAGE PROCESSING}
\label{appBsec} 

In this appendix we process HST infrared imaging of B1608+656 obtained with the 
NIC1 camera through the filter F160W.
The observations were conducted in 1997 February as part of HST proposal 7422.
The NIC1 detector is an array of 256x256 pixels whose elements 
are divided into four independent quadrants of 128x128 pixels. 
The pixel size is 0.043''
and the conversion from pixel 
data in DN = counts/sec into electrons/sec is determined 
by the factor of Gain=5e$^{-}$DN$^{-1}$. 
There are 7 exposures with F160W, designated h1, h2.. h7
in chronological order.
The first five exposures are 3839.94 seconds long, h6 is 2047.86 seconds long,
and h7 is 895.92 seconds long. 

The images suffer from some typical anomalies of NIC1. 
All frames present the NIC1 high noise region in the upper left quadrant.
This region has a sensitivity that is factor of 2-3 lower than the mean of the array
and its noise gets highly amplified during the calibration process. 
The lens is located in the lower right quadrant, but is still 
marginally affected by the high noise region.
The second anomaly present in all frames is the bad central row,
which affects all pixels in row 128 of the array
and sometimes the adjacent rows as well.   
The lens extends over row 128 in frames h3 and h6.
We therefore discard rows from 126 to 130 in those frames.
We also discard rows 127 to 129 that can marginally affect the area around the lens 
in all frames.
There is also evidence for quadrant dependence of the signal, 
which is not unexpected since each quadrant has its own amplifier.
This can be relevant in h3 and h6 where the lens extends over 
two quadrants.
Frame h6 does not seem to be affected, but the signal in the north 
right quadrant of h3 
has higher amplification than the rest of the frame and the data in that
quadrant will not be included. 
The frames also show evidence of glow or
maximums in the corner of each quadrant, due to the readout amplifiers,
but that doesn't seem to affect the lens area.

The sky is very noisy, its level differs from frame to frame,
and it also varies across each frame.
With the high noise region of NIC1 located in the north-left quadrant, 
and B1608+656 in the lower-right quadrant, the noise creates
a sky level in the lens area that can be modeled as  
a gradient at -45$^{\rm o}$ direction (degrees counterclockwise from +y pixel 
coordinates). The noise level follows the same pattern.
We first determine the mean sky level 
in the lower-right quadrant of NIC1 and subtract it.
These values are listed in the second column of Tab.~\ref{hdata}.
Then we model the remaining sky gradient 
as ${\rm C (x-y-255)^2}$, 
where x and y are pixel coordinates and C is a constant.
The fitted values of C are listed in Tab.~\ref{hdata}.

%%%%%%%%%%%%%%%%%%%%%%%%%%%%%%%%%%%%%%%%%%%%%%%%%%%%%%%%%%%%
\begin{deluxetable}{ccccc}
\tablecolumns{5}
\tablewidth{0pc}
\tablecaption{Sky Level, Shift, and Weight Factor in H-band Frames \label{hdata}}
\tablehead{
\colhead{Frame} & \colhead{Sky Level\tablenotemark{a}} & 
\colhead{Sky Gradient C\tablenotemark{a,b}} 
& \colhead{Pixel Shift\tablenotemark{c}}& \colhead{Weight Factor}}
\startdata
h1 & 714  $\pm$ 242  & 0.0028  & 22.91 ~~ 22.78 & 0.31 \\
h2 & 402  $\pm$ 136  & 0.0017  & -0.10 ~~ 22.74 & 0.56 \\
h3 & 96.7 $\pm$ 88.2 & 0.0001  & 22.95 ~~ -0.32 & 0.86 \\
h4 & 154  $\pm$ 76.0 & -0.0002 & 46.14 ~~ 23.09 & 1.0 \\
h5 & 96.2 $\pm$ 87.1 & -0.0022 & 23.08 ~~ 46.42 & 0.87 \\
h6 & 38.3 $\pm$ 46.0 & -0.0013 & -0.24 ~~ -0.23 & 0.87 \\
h7 & 10.7 $\pm$ 31.1 & -0.0002 & 46.24 ~~ 46.52 & 0.56 \\
\enddata
\tablenotetext{a}{Sky measured in units of photons.} 
\tablenotetext{b}{After subtracting the constant sky level,
the remaining sky gradient at each frame is modeled as 
${\rm C(x-y-255)^2}$, with $\left(x,y \right)$ pixel coordinates.}
\tablenotetext{c}{The two values correspond to x and y shifts 
in the pixel grid.}
\end{deluxetable} 
%%%%%%%%%%%%%%%%%%%%%%%%%%%%%%%%%%%%%%%%%%%%%%%%%%%%%%%%%%%%

Using the centroids of A, B, C, D and G1 we determine the relative
shifts between frames, but we don't apply them yet (the values 
are listed in Tab.~\ref{hdata}).
H frames need a further rotation, translation and magnification
to be registered to the V and I pixel grid.
At this point it is better to register the H frames directly to the V and I 
grid before the combination. This involves only one geometrical
transformation, while shifting, combining and 
registering the final H image to V and I involves two.
Using A and B centroids we determine a rotation of 
7.903$^{\rm o}$ (degrees counterclockwise)
is needed to register H frames to V and I.
A magnification factor of 1.05819 is also required
because H frames have 0.043'' pixels while V and I frames 
have 0.0455'' pixels.
Combining all this information
we determine the transformation
needed to register each H frame to the V and I grid of pixel coordinates.

\begin{figure}
	\figurenum{3}
       \epsscale{0.82}
%        \plotone{/home/surpi/plot/ploth.ps}
        \plotone{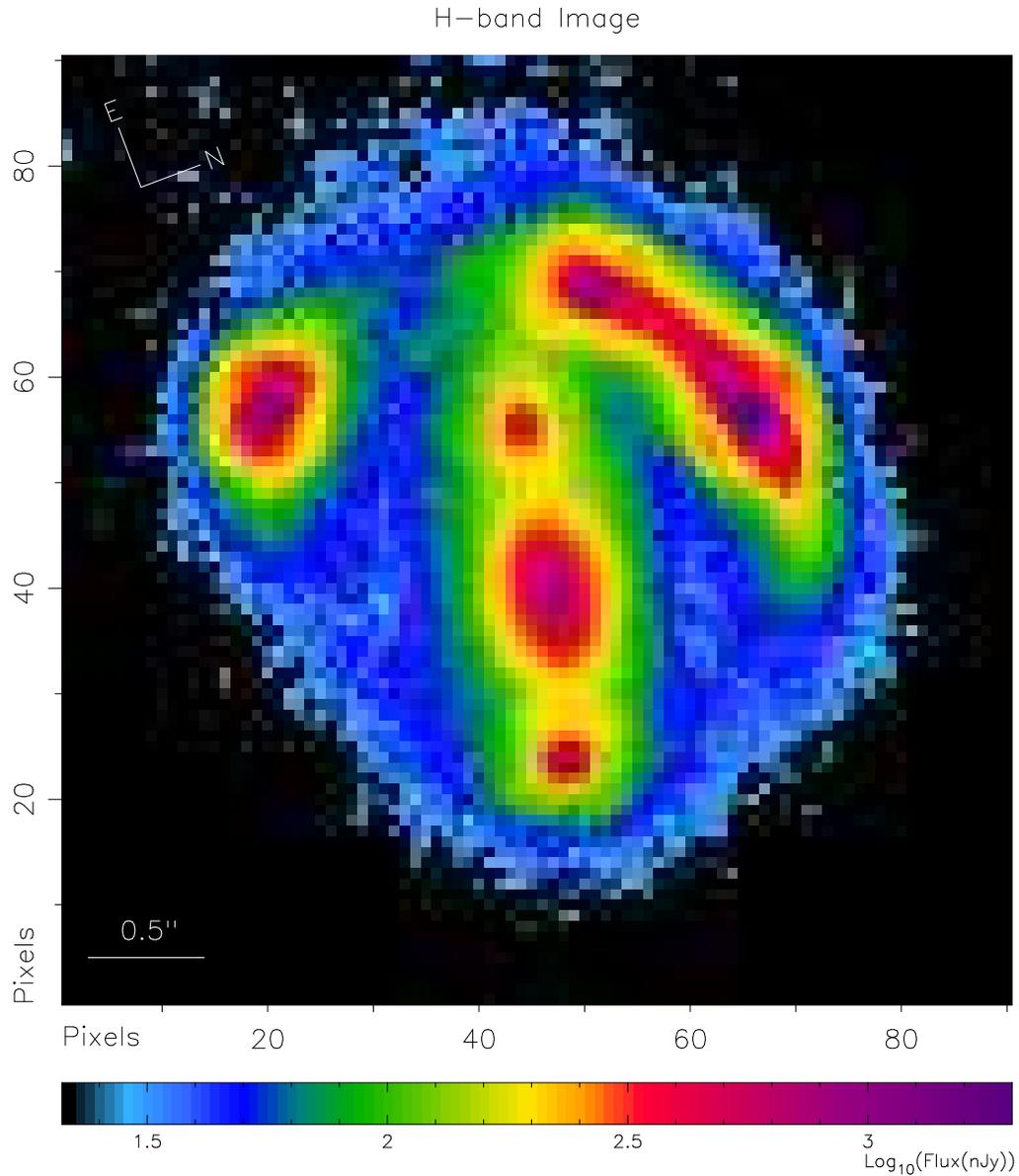}
       \caption{H band image of B1608+656 from F160W exposures by Chris Fassnacht 
(HST proposal 7422). 
Image is the average combination of 7 frames with a total exposure time of 
6 hrs 9 min. The ring like emission at radius of approximately 
4 pixels around each of the
multiple images is artificial and caused 
by diffraction at 16000 \AA~ in the 2.4m diameter camera.
Only fluxes above 3 sigma of the sky noise are displayed.}
       \label{ploth}
\end{figure}

\begin{figure}
	\figurenum{4}
       \epsscale{0.33}
%        \plotone{/home/surpi/plot/plotstn.ps}
        \plotone{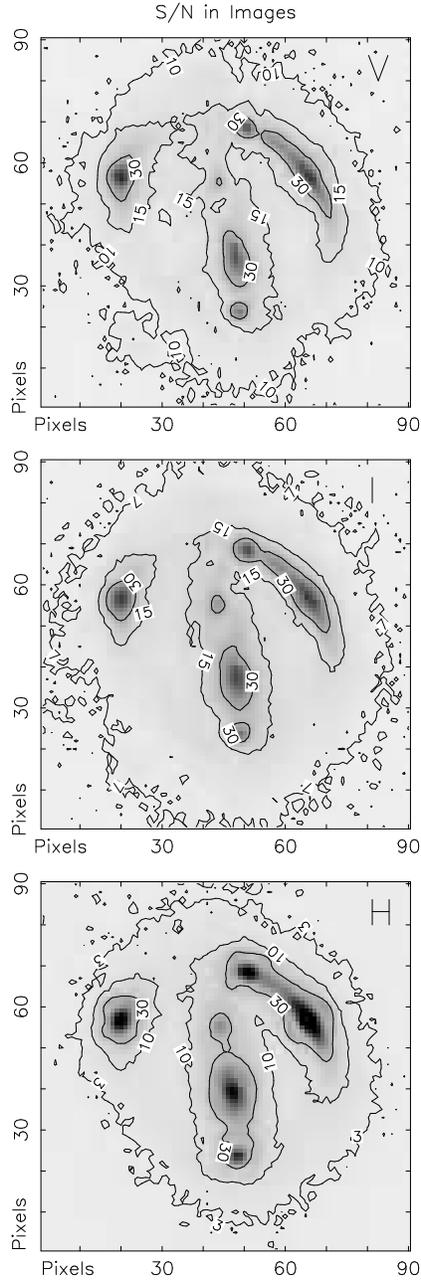}
       \caption{Contours of constant signal-to-noise in the V, I, and H band images.
The emission from the lens has approximately
the same S/N in V and I bands. In the low emission region, the S/N of H 
is lower than that in V and I due to the high noise region in NIC1.
Displayed contours of S/N~$\sim$~10, 7, and 3 in V, I, and H respectively 
correspond to emission at 3$\sigma$ level. 
Contours of S/N~$\sim$~30 correspond approximately
to isophotes enclosing 50\% of the source flux. Peak values of S/N are
90, 85, and 200 for V, I, and H bands.}
       \label{plotstn}
\end{figure}

\begin{figure}
	\figurenum{7}
       \epsscale{0.4}
%        \plotone{/home/surpi/plot/plotdec.ps}
        \plotone{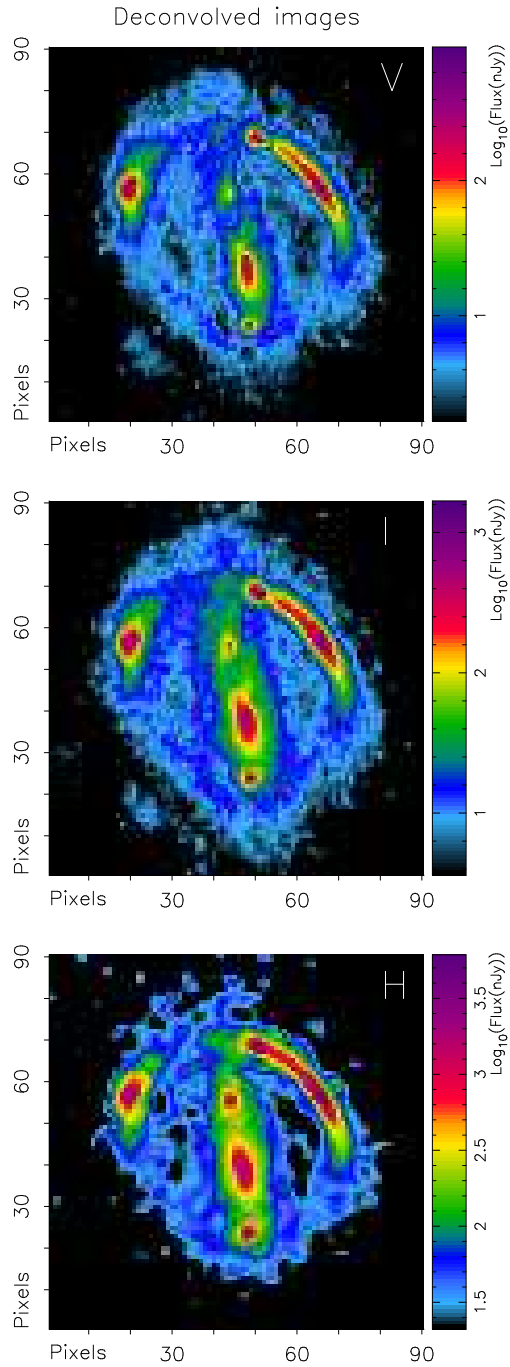}
       \caption{Images deconvolved with Lucy-Richardson method. 
The color scale used at each band
is the same than in Fig.~\ref{plotv}, \ref{ploti}, and \ref{ploth}. 
The deconvolution 
has peaked the core fluxes of the source by factors~$\sim$
2, 2 and 3 at V, I, and H respectively. H image still shows residuals
of the airy rings around images A, C, and D.}
       \label{plotdec}
\end{figure}

There is no contamination with cosmic rays, 
as they have been filtered out by use of MULTIACCUM mode.
Only a few bad single pixels need to be rejected. 
We average the frames using weights to account their different
exposure time and background noise error, the weight factors used 
are listed in Tab.~\ref{hdata}.
The resulting H image is shown in Fig.~\ref{ploth}.
The S/N in the image is hard to characterize.
The statistical properties of the noise can be affected
by the rotation and re-scaling performed to the individual frames.
Besides, even before the registration, the difference between the 
background noise in the frames makes the general formula 
used to estimate the S/N for V and I bands not appropriate. 
We construct here a S/N map estimating the noise from the standard deviation 
in the data at each pixel, appropriately smoothed. Contours of S/N=3,10 and 30
are shown in Fig.~\ref{plotstn}.

\bibliography{gravlens}

\vfill\eject

%%%%%%%%%%%%%%%%%%%%% Figure 1 %%%%%%%%%%%%%%%%%%%%%%%%%%%%%%%%%%%%%%
%%%%%%%%%%%%%%%%%%%%%% Figure 2 %%%%%%%%%%%%%%%%%%%%%%%%%%%%%%%%%%%%%%
%%%%%%%%%%%%%%%%%%%%%% Figure 3 %%%%%%%%%%%%%%%%%%%%%%%%%%%%%%%%%%%%%%
%%%%%%%%%%%%%%%%%%%%%% Figure 4 %%%%%%%%%%%%%%%%%%%%%%%%%%%%%%%%%%%%%%
%%%%%%%%%%%%%%%%%%%%%% Figure 5 %%%%%%%%%%%%%%%%%%%%%%%%%%%%%%%%%%%%%%
%%%%%%%%%%%%%%%%%%%%%% Figure 6 %%%%%%%%%%%%%%%%%%%%%%%%%%%%%%%%%%%%%%
%%%%%%%%%%%%%%%%%%%%%%%%% FIGURE 7
%%%%%%%%%%%%%%%%%%%%%%%%% FIGURE 8
%%%%%%%%%%%%%%%%%%%%%%%%%%%%%%%%%%%%%%%%%%%%%%%%%%%%%%%%%%%%%%%%%%%%%%%%%%%%%%%

\end{document}